\documentclass[oldversion]{aa} 
\usepackage{graphicx}
\usepackage{txfonts,color}
\usepackage[authoryear]{natbib}

\bibpunct{(}{)}{;}{a}{}{,} 
\begin{document}
\newcommand{\mps}{\,m\,s$^{-1}$}
\newcommand{\kms}{\,km\,s$^{-1}$}
\newcommand{\msec}{\,m\,s$^{-1}$}
\newcommand{\ha}{H$\alpha$ }

   \title{Chromospheric features of LQ Hydrae from \ha line profiles}

\author{A.~Frasca\inst{1}
\and Zs.~K\H{o}v\'{a}ri\inst{2}
\and K.~G.~Strassmeier\inst{3}
\and K.~Biazzo\inst{1}
}
   \offprints{A.~Frasca}

   \institute{INAF -- Catania Astrophysical Observatory, via S. Sofia 78, I-95123, Catania, Italy\\
    \email{antonio.frasca@oact.inaf.it, katia.biazzo@oact.inaf.it}
    \and
    Konkoly Observatory, H-1525 Budapest, P.O.Box 67, Hungary\\
             \email{kovari@konkoly.hu}
         \and
             Astrophysical Institute Potsdam, An der Sternwarte 16, D-14482 Potsdam, Germany\\
             \email{kstrassmeier@aip.de}
}

\date{Received 14 Nov 2007 / Accepted 3 Jan 2008}
\abstract{We analyze the \ha spectral variability of the
rapidly-rotating K1-dwarf LQ\,Hya using high-resolution \ha spectra
recorded during April-May 2000. Chromospheric parameters were
computed from the \ha profile as a function of rotational phase. We
find that all these parameters vary in phase, with a higher
chromospheric electron density coinciding with the maximum \ha emission. 
We find a clear rotational modulation of the \ha emission
that is better emphasized by subtracting a reference photospheric template
built up with a spectrum of a non-active star of the same spectral
type.
A geometrical plage model applied to the \ha variation curve allows
us to derive the location of the active regions that come out to be
close in longitude to the most pronounced photospheric spots found with Doppler imaging
applied to the photospheric lines in the same spectra. Our analysis
suggests that the \ha features observed in LQ~Hya in 2000 are a
scaled-up version of the solar plages as regards dimensions and/or
flux contrast. No clear indication of chromospheric mass motions
emerges.


\titlerunning{Chromospheric features of LQ Hya from \ha line profiles}
}
   \maketitle


\section{Introduction}\label{intro}

According to the solar-stellar analogy, chromospheric stellar activity can be
established by the presence of emission in the  core of the Balmer \ha line.
As in the Sun, \ha emission intensification has often been observed
in surface features (plages) spatially connected with the photospheric starspots
\citep[see, e.g.,][and references therein]{Cata00,Bia06,Bia07}.
Thus, the time variability of the \ha spectral features can be used
to estimate the basic properties of the emitting sources, allowing the geometry of 
the stellar chromosphere to be mapped.

In this paper we analyze the \ha spectral variability of the young,
rapidly-rotating ($P_{\rm rot}\approx 1.6$\,d) single K2-dwarf
\object{LQ\,Hydrae} (HD\,82558 = Gl\,355) using 15 high-resolution
spectra recorded during April-May 2000. All \ha spectra were
acquired simultaneously with the mapping lines used for the year-2000 
Doppler imaging study presented in our previous paper \citep[][hereafter
Paper~I]{Kova04}.
LQ\,Hya was first recognized as  a chromospherically active star
through \ion{Ca}{ii} H\&K emission \citep{bide81,hein81}.
The \ha absorption filled in by chromospheric emission was reported first by \citet{feke86},
and LQ\,Hya was classified as a BY\,Dra-type spotted star.
Variable \ha emission-peak asymmetry was investigated by \citet[][hereafter Paper~II]{Stra93},
who attributed it to the presence of chromospheric velocity fields in the \ha forming layer
probably surrounding photospheric spots.

In our spectroscopic study in Paper~I, we presented Doppler images
using the \ion{Fe}{i}-6411, \ion{Fe}{i}-6430, and \ion{Ca}{i}-6439
lines for both the late April data and the early May 2000 dataset.
Doppler imaging was supported by simultaneous photometric
measurements in Johnson-Cousins $VI$ bands. Doppler images showed
spot activity uniformly at latitudes between $-20$\degr and
$+50$\degr, sometimes with high-latitude appendages, but without a
polar spot. Comparing the respective maps from two weeks apart,
rapid spot evolution was detected, which was attributed to strong
cross-talks between the neighboring surface features through
magnetic reconnections. In Table~\ref{chart} we give a summary of
the stellar parameters as they emerged from Paper~I.

\begin{table}
\caption{Astrophysical data for LQ\,Hya \citep[adopted from][]{Kova04}}
\label{chart}
\centering
 \begin{tabular}{ll}
  \hline\hline
  \noalign{\smallskip}
  Parameter                & Value \\
  \noalign{\smallskip}
  \hline
  \noalign{\smallskip}
  Classification           & K2 V \\
  Distance (Hipparcos)     & 18.35$\pm$0.35 pc \\
  $(B-V)_{\rm Hipparcos}$  & 0.933$\pm$ 0.021 mag \\
  $(V-I)_{\rm Hipparcos}$  & 1.04$\pm$ 0.02 mag \\
  Luminosity, $L$          & 0.270$\pm$0.009~L$_\odot$ \\
  $\log g$                 & 4.0$\pm$0.5 \\
  $T_{\rm eff}$            & 5070$\pm$100 K \\
  $v\sin i$                & 28.0$\pm$1.0 km/s \\
  Inclination, $i$         & 65$^{\circ}$$\pm$10$^{\circ}$ \\
  Period, $P_{\rm rot}$    & 1.60066$\pm$0.00013 days \\
  Radius, $R$              & 0.97$\pm$0.07 $R_\odot$ \\
  Mass                     & $\approx$ 0.8 $M_\odot$ \\
  Age                      & $\approx$ ZAMS \\
  \noalign{\smallskip}
  \hline
 \end{tabular}
\end{table}

The observations are again briefly presented in Sect.~\ref{data} and
the method used for the \ha spectral study is described in
Sect.~\ref{sec:analysis}. The results are presented in
Sect.~\ref{results}. Since the \ha observations of this paper were
included in the spectroscopic data used in Paper~I to reconstruct
the year-2000 Doppler images, in Sect.~\ref{disc} we take the
opportunity to compare the photospheric features with the contemporaneous 
H$\alpha$-emitting regions.

\section{Spectroscopic observations}
\label{data}

 The series of 15 \ha spectra was collected, one spectrum per night, during two observing runs
(April 4--9, 2000, and April 25--May 3, 2000) at the Kitt Peak National
Observatory (KPNO) with
the 0.9 m coud\'e-feed telescope. The 3096$\times$1024 F3KB CCD
detector was employed, together with grating A, camera 5, and the long collimator.
The spectra were centered at 6500\,\AA\ with a wavelength range of 300\,\AA.
The effective resolution was $\approx$\,28\,000 (11\,\kms).
 We achieved signal-to-noise ratios (S/N) of about 250 in 45-min integration time,
with the only exception being the spectrum at 2.726 phase for which an S/N of just 50 
could be reached due to bad sky conditions.

The mean HJDs of the observations, the phases, as well as the radial
velocities, are summarized in Table~\ref{T1}. For phasing the spectra,
$HJD = 2448270.0 + 1.600656 \times E$  was used, where the
arbitrarily chosen zero point is the same as used for our former
Doppler imaging studies \citep{Stra93,Kova04}. As radial velocity
(RV) standard star, $\beta$\,Gem ($v_{\rm r}=3.23$\kms) was
measured, except for HJD 2451643.744 when 16\,Vir ($v_{\rm r}=36.48$\kms) was 
observed \citep{scar}. For more details we refer
to Paper~I.

\begin{table}
\caption{Observing log and radial velocities }
\label{T1}
\begin{flushleft}
 \begin{tabular}{lrll}
  \hline\hline
  \noalign{\smallskip}
$HJD$   & Phase   & $v_{\rm r}$   & $\sigma_{v_{\rm r}}$  \\
        &         & (\kms)	    & (\kms)	 \\
  \hline
2451639.681 &    0.187  & 7.1 &       0.6 \\
2451640.647 &    0.791  & 7.8 &   0.6 \\
2451641.648 &    1.416  & 8.6 &   0.5 \\
2451642.662 &    2.050  & 7.4 &   0.6 \\
2451643.744 &    2.726  & 7.6$^a$ &   2.4 \\
2451644.677 &    3.309  & 8.3 &   0.6 \\
2451660.683 &   13.308  & 8.8 &   0.5 \\
2451661.668 &   13.924  & 8.7 &   0.6 \\
2451662.652 &   14.538  & 9.5 &   0.6 \\
2451663.654 &   15.165  & 7.5 &   0.5 \\
2451664.667 &   15.798  & 9.0 &   0.6 \\
2451665.653 &   16.413  & 9.8 &   0.5 \\
2451666.661 &   17.043  & 8.0 &   0.5 \\
2451667.665 &   17.670  & 7.4 &   0.6 \\
2451668.670 &   18.298  & 8.1 &   0.7 \\
  \noalign{\smallskip	       }
  \hline
 \end{tabular}\\
\end{flushleft}

$^a$ with 16\,Vir as RV standard
\end{table}

\section{Data analysis}
\label{sec:analysis}

The Balmer \ha line is  a useful and easily accessible
indicator of chromospheric activity in the optical spectrum. It has
been proven to be very effective for detecting chromospheric plages
both in the Sun and in active stars, due to its high contrast
against the surrounding chromosphere \citep[see, e.g.,][]{Fra98,Bia06,Bia07}.

However, the \ha line is formed in a wide depth range in the stellar
atmosphere, ranging from the temperature minimum to the upper
chromosphere. To extract the chromospheric contribution
from the line core, we apply  a method usually called ``spectral synthesis", 
which consists in the subtraction of synthetic spectra or of observed spectra
of non-active standard stars (reference spectra) \citep[see,
e.g.,][]{Herb85, Bar85, Fra94, Montes95}. The difference between
observed and reference spectra provides  the net \ha emission, 
which can be integrated to estimate the total radiative chromospheric loss 
in the line.

We point out that one can obtain the true chromospheric \ha emission with this 
method only if the chromosphere is also optically thin inside the active 
regions or if the plages are not extended enough to appreciably affect the 
photospheric spectrum. 
In the opposite case, the subtraction of the underlying photospheric
profile would overestimate the \ha chromospheric emission.

\begin{figure}[tbh]
\includegraphics[width=\columnwidth]{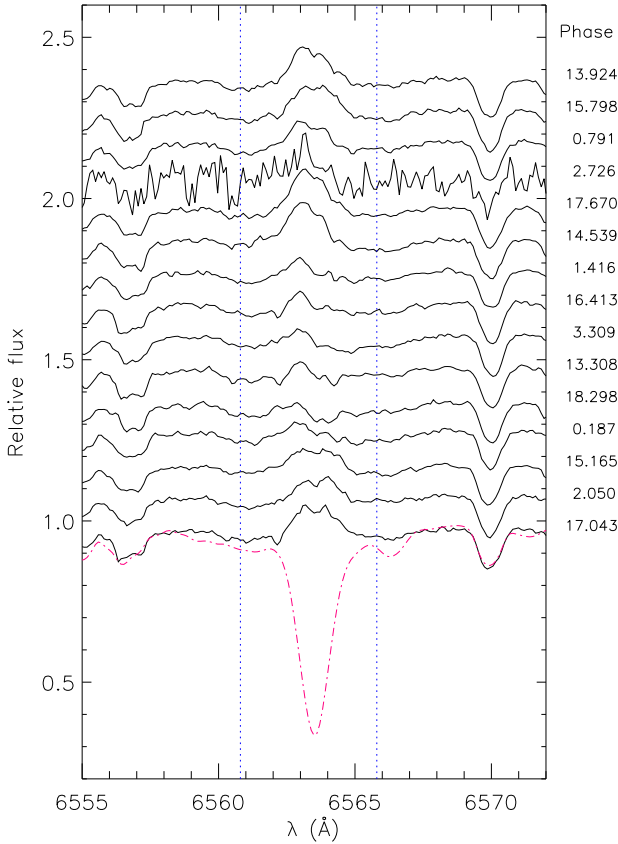}     
\caption{Time series of \ha profiles for LQ~Hya in April-May 2000 sorted in phase order. The non-active template
built up with a spectrum of the non-active, slowly-rotating K2\,V star HD\,3765 is shown by a  dash-dotted line 
 at the bottom. The vertical dotted lines mark the integration interval for the {\it raw} \ha equivalent width
($EW_{\rm H\alpha}$).}
\label{fig:spectra}
\end{figure}

For this reason we have  also measured the \ha equivalent width in
the observed spectra ($EW_{\rm H\alpha}$), fixing a constant
integration window of 5.0\,\AA, i.e. integrating the emission
core inside the absorption wings (see Fig.~\ref{fig:spectra}).	
The net \ha equivalent width, $\Delta EW_{\rm H\alpha}$, 
was instead measured in the residual spectra obtained from the
original ones with the ``spectral synthesis" method.
The non-active template was built using a
high-resolution ($R=42\,000$) spectrum of the K2\,V star
\object{HD\,3765} retrieved from the ELODIE archive. We chose
HD\,3765 because its spectral type and color index $B-V=0.942$
\citep{Weis96} are nearly identical with those of LQ\,Hya, because its 
rotation velocity is very low \citep[$v\sin i=2.5$\kms,][]{Stra00}, and because it has a low level 
of magnetic activity compared to LQ\,Hya.
The spectrum was first degraded to the resolution of the KPNO spectra 
by convolving it with a  Gaussian kernel and subsequently broadened by convolution
with a rotational profile corresponding to the $v\sin i$ of 28\kms
of LQ\,Hya.

The error of the equivalent width, $\sigma_{\rm EW}$, was evaluated by multiplying the 
integration range by the photometric error on each point. This was estimated by the standard 
deviation of the observed flux values on the difference spectra in two spectral regions near 
the H$\alpha$ line. The values of $EW_{\rm H\alpha}$, $\Delta EW_{\rm H\alpha}$, and $\sigma_{\rm EW}$
are reported in Table~\ref{T2}.

We also measured the central wavelength of the \ha emission both in the raw (observed) and 
residual spectra by evaluating the centroid of the core emission. This quantity provided us with 
the radial velocity shifts between the \ha emission and the photosphere measured both in the raw
($\Delta V_{\rm em}$) and in the residual spectra ($\Delta V_{\rm em}^{\rm res}$). The values of 
$\Delta V_{\rm em}$ and $\Delta V_{\rm em}^{\rm res}$  are reported in Table~\ref{T2} as well.

\section{Results}\label{results}

In  the LQ\,Hya spectra, the \ha line is a strongly variable feature displaying absorption wings 
and a core changing from a completely filled-in configuration to a moderate emission above the continuum 
(see Fig.~\ref{fig:spectra}).
A faint single peak just filling in the line core, without reaching
the continuum, is observed at the phases of minimum \ha emission
($\phi$=0.2$-$0.4), whereas double-peaked emission profiles are
observed with a central reversal at the other phases. The ``blue''
emission peak is often stronger than the ``red'' one, although at
some phases, nearly equal peak heights and even reversed intensity
ratios are also observed. The \ha profiles are shown in
Fig.~\ref{fig:spectra} along with the photospheric template built up
with an HD\,3765 spectrum (see Sect.~\ref{sec:analysis}).

\begin{figure}[tbh]
\includegraphics[width=\columnwidth]{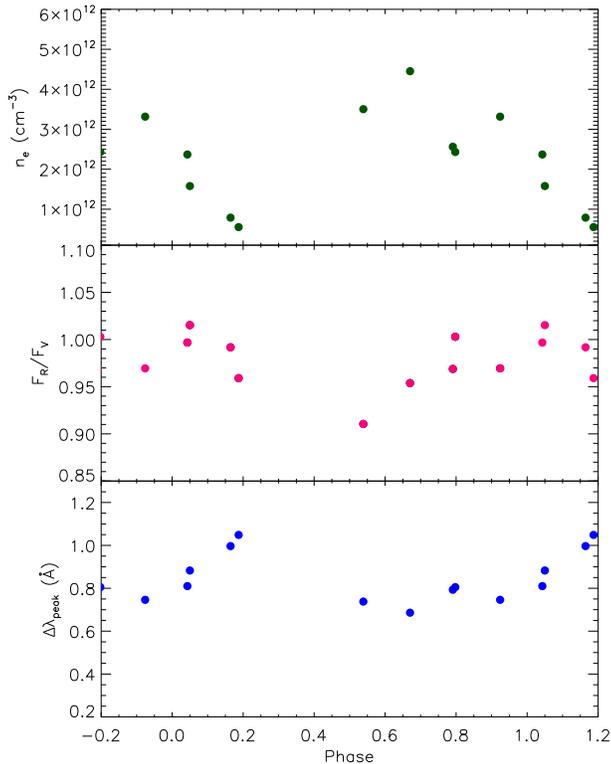}
\caption{{\it From top to bottom}. The computed electron density
$n_{\rm e}$ (cm$^{-3}$), the flux ratio between the red and the blue
peaks, and the peak separation in \AA\  against the rotational
phase.} \label{fig:Ne}
\end{figure}

We estimated the chromospheric electron density in LQ\,Hya adopting the assumptions from Paper\,II, i.e. an
isothermal chromosphere with $T_{\rm chrom}=10\,000$\,K for which the optical depth in the \ha center can be
derived by the formulation of \citet{Cram79}:
\begin{equation}
\ln\tau_{\rm chrom} = \left(\frac{\Delta\lambda_{\rm peak}}{2\Delta\lambda_{\rm D}}\right)^2,
\end{equation}
where $\Delta\lambda_{\rm peak}$ is the wavelength separation of the
blue and red emission peaks, and $\Delta\lambda_{\rm D}=0.28$\,\AA\
is the chromospheric Doppler width. We measured
$\Delta\lambda_{\rm peak}$ on the observed spectra by fitting two
Lorentzian functions to all the double-peaked \ha profiles.
Obviously, this analysis could not be applied to the single-peaked
spectra close to the minimum emission at phase 0.2--0.5.  Combining equations~8 and 
14 in \citet{Cram79}, the electron density $n_{\rm e}$ in cm$^{-3}$ can be evaluated as
\begin{equation}
n_{\rm e} = 1.67\cdot 10^{14}\frac{F_{\rm max}}{F_{\rm cont}}\frac{B(T_{\rm eff})}{B(T_{\rm chrom})}\tau_{\rm chrom}^{-1},
\end{equation}
where $\frac{F_{\rm max}}{F_{\rm cont}}$ is the ratio of the  \ha peak to
the  continuum flux, as measured on our spectra, and $B$ is the Planck function calculated for the
effective temperature of LQ\,Hya ($T_{\rm eff}=5100$\,K) and at
$T_{\rm chrom}=10\,000$\,K. The optical depth at the \ha center
changes from about 30 to 4 and the electron density from $5\cdot
10^{11}$  to $4\cdot 10^{12}$ cm$^{-3}$ from phase $\approx 0.2$
(near the minimum \ha emission) to the phases of maximum emission
($\phi$=0.6$-$0.8, see Fig.\,\ref{fig:Ne}). In the same figure, the
flux ratio between the red and the blue peaks $\frac{F_{\rm
R}}{F_{\rm V}}$ and the peak separation $\Delta\lambda_{\rm peak}$
are plotted versus the rotational phase. All these quantities appear
modulated by the stellar rotation.  Symmetric profiles 
($\frac{F_{\rm R}}{F_{\rm V}}\simeq 1$) tend to be observed when the 
most active regions, judging from the photospheric spot contrast, are 
in the visible hemisphere, similar to what was found
by Strassmeier et al. in Paper~II.

\begin{figure}[tbh]
\includegraphics[width=\columnwidth]{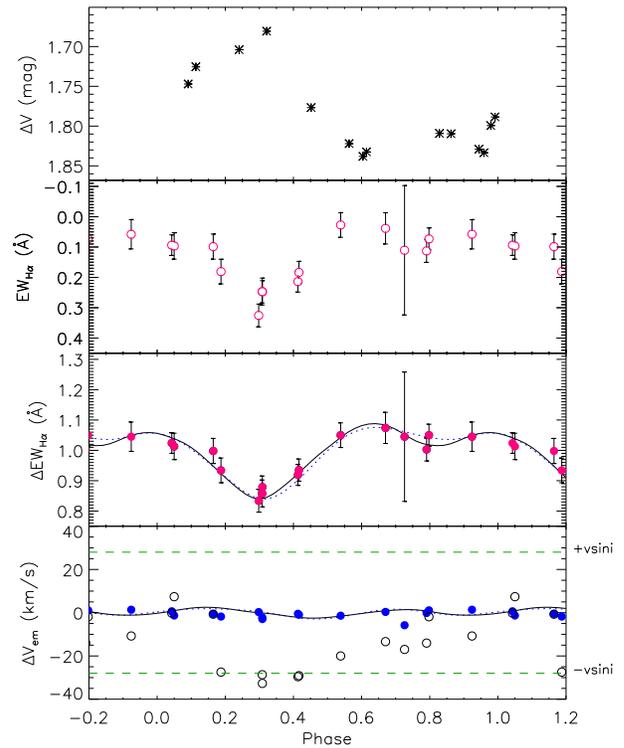}
\caption{{\it Top panel}:  $V$ light curve contemporaneous to the spectra. {\it Middle panels}: Rotational 
modulation of the {\it raw} \ha equivalent width ($EW_{\rm H\alpha}$, open circles) and of the {\it residual} 
\ha EW ($\Delta EW_{\rm H\alpha}$, filled circles). 
{\it Bottom panel}: Velocity shift between \ha emission and photospheric lines both for observed (open circles) 
and residual spectra (filled circles).
The continuous lines in the middle and bottom panels represent the best fit of our 3-plage model
(cf. Fig.~\ref{fig:palle}), while the dotted lines represent the result of our 2-plage model.}
\label{fig:rotmod}
\end{figure}

 The \ha equivalent width measured in the observed ($EW_{\rm H\alpha}$) and 
in the residual spectra ($\Delta EW_{\rm H\alpha}$), the radial velocity shifts
of the \ha emission ($\Delta V_{\rm em}$ and $\Delta V_{\rm em}^{\rm res}$), and 
the contemporaneous light curve in the $V$ band (from Paper~I) are plotted in Fig.~\ref{fig:rotmod} 
as functions of the rotational phase. A remarkable anti-correlation of $V$ brightness
and \ha emission is apparent.

The highest electron density is observed when the \ha line is
strongest, indicating a denser chromosphere above more extended
active regions, the opposite of what was observed in 1991 (Paper~II).
 We also found an average value for the chromospheric electron
density that was higher than the one reported in Paper~II ($2\cdot
10^{11}$\,cm$^{-3}$). Moreover, the rotational modulation of all the
quantities deduced by the \ha line analysis is much more evident in
the present data.  Presumably, in April-May 2000, LQ\,Hya
was more active than in 1991, and it displayed more and larger active
regions.

%

\begin{table}
\caption{Equivalent widths of the {\it raw} ($EW_{\rm H\alpha}$) and the {\it residual} spectra
($\Delta EW_{\rm H\alpha}$),  errors of the equivalent widths ($\sigma_{\rm EW}$), and 		
velocity shifts $\Delta V_{\rm em}$ and $\Delta V_{\rm em}^{\rm res}$ for the {\it raw} and the {\it residual} spectra,
respectively.}
\label{T2}
\centering
 \begin{tabular}{lcccrr}
  \hline\hline
  \noalign{\smallskip}
 Phase   & $EW_{\rm H\alpha}$ & $\Delta EW_{\rm H\alpha}$ & $\sigma_{\rm EW}$ & $\Delta V_{\rm em}$ & $\Delta V_{\rm em}^{\rm res}$ \\     \\
         &  (\AA)             &  (\AA)            &  (\AA)      &(\kms)&(\kms)\\
  \hline
   0.187 &  0.172   &	0.934  &    0.041  &  $-27.46$  &   $-1.78$  \\
   0.791 &  0.117   &	1.003  &    0.038  &  $-14.05$  &   $-0.13$  \\
   1.416 &  0.183   &	0.935  &    0.036  &  $-29.11$  &   $-0.96$  \\
   2.050 &  0.107   &	1.013  &    0.043  &  $  7.40$  &   $-1.29$  \\
   2.726 &  0.116   &	1.045  &    0.213  &  $-17.00$  &   $-5.80$  \\
   3.309 &  0.239   &	0.879  &    0.037  &  $-32.70$  &   $-2.92$  \\
  13.308 &  0.248   &	0.858  &    0.044  &  $-28.68$  &   $-2.19$  \\
  13.924 &  0.058   &	1.045  &    0.048  &  $-10.77$  &   $ 1.38$  \\
  14.538 &  0.025   &	1.050  &    0.041  &  $-20.01$  &   $-1.38$  \\
  15.165 &  0.099   &	0.998  &    0.041  &  $ -0.69$  &   $-0.56$  \\
  15.798 &  0.073   &	1.050  &    0.036  &  $ -1.87$  &   $ 1.09$  \\
  16.413 &  0.215   &	0.919  &    0.035  &  $-29.60$  &   $-0.49$  \\
  17.043 &  0.094   &	1.024  &    0.034  &  $ -0.18$  &   $ 0.65$  \\
  17.670 &  0.038   &	1.074  &    0.051  &  $-13.38$  &   $ 0.36$  \\
  18.298 &  0.325   &	0.834  &    0.038  &  $-43.07$  &   $ 0.29$  \\
  \noalign{\smallskip}
  \hline
 \end{tabular}\\
\begin{flushleft}
\end{flushleft}
\end{table}

\section{Discussion and conclusions}\label{disc}

To obtain information about the surface location of the
active regions in the chromosphere of LQ~Hya, we applied a simple
geometric {\it plage} model to the rotationally modulated
chromospheric emission. This method had been described and
successfully applied to the \ha modulation curves of several other
active stars \citep[e.g.,][]{Fra00, Bia06, Fra07}. On LQ~Hya, two
circular bright spots (plages) are normally sufficient for reproducing
the observed variations within the data errors
\citep[cf.][]{Fra00,KoBa97}. In our model the flux ratio between
plages and the surrounding chromosphere ($F_{\rm plage}/F_{\rm
chrom}$) is a free parameter. Solar values of $F_{\rm
plage}/F_{\rm chrom}\,\approx\,2$, as deduced from averaging many
plages in H$\alpha$  \citep[e.g., ][]{Elli52,LaBo86,Ayres86}, are too
low to model the high amplitudes of H$\alpha$ $\Delta EW$ curves
in very active stars \citep{Bia06,Fra07}. In fact, extremely large
plages, covering a significant fraction of the stellar surface,
would be required with such a low flux ratio, and they could not
reproduce the observed modulation.
In order to achieve a good fit, a flux ratio of 5, which is also
typical of the brightest parts of solar plages or of flare regions
 \citep[e.g., ][]{Sves76,Ziri88},
was adopted.

The solutions essentially provide the longitude of the plages 
and give only rough estimates of their latitude and size. 
 The information about the latitude of surface features can be recovered
from the analysis of spectral-line profiles broadened by the stellar
rotation, as we did for the spots of LQ~Hya with Doppler imaging in Paper I.
A similar technique cannot reach the same level of accuracy when applied to 
the \ha profile whose broadening is dominated by chromospheric heating effects,
which are particularly efficient in the active regions \citep[e.g.,][]{Lanzafame00}.

We searched for the best solution by varying the longitudes, latitudes,
and radii of the active regions. The radii are strongly dependent on
the assumed flux contrast $F_{\rm plage}/F_{\rm chrom}$. Thus, only
the combined information between plage dimensions and flux contrast,
i.e. some kind of plage ``luminosity'' in units of the quiet
chromosphere ($L_{\rm plage}/L_{\rm quiet}$), can be deduced as a
meaningful parameter. Note also that we cannot estimate the true
quiet chromospheric contribution (network), since the H$\alpha$
minimum value, $\Delta EW_{\rm quiet}=0.84$\,\AA, could still be
affected by a homogeneous distribution of smaller plages. However,
such an approximation seems valid for LQ~Hya because our aim was
only to compare the spatial distribution of the main surface
inhomogeneities at chromospheric and photospheric levels.

We also searched for solutions with three active regions,
finding a small but possibly significant increase in the goodness of
the fit, with the $\chi^2$ passing from 5.77 to 4.45 (see
Fig.~\ref{fig:rotmod}). The model with only two plages requires
features at rather high latitudes (Table~\ref{tab:param_plages}) 
to explain the nearly flat and long-lasting maximum, while a
3-plage model allows placing plages at lower latitudes, in better
agreement with the spot locations from Doppler imaging.
 However, the longitudes of the two largest plages do not
change very much (less than 15 degrees).

\begin{figure}[tbh]
\vspace{0.5cm}
\begin{center}
\includegraphics[width=2.5cm]{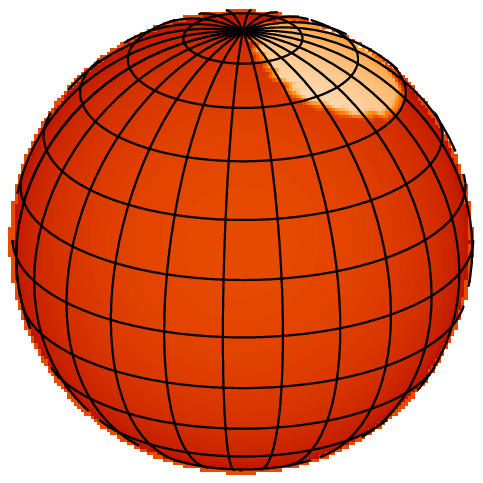}	
\hspace{0.4cm}
\includegraphics[width=2.5cm]{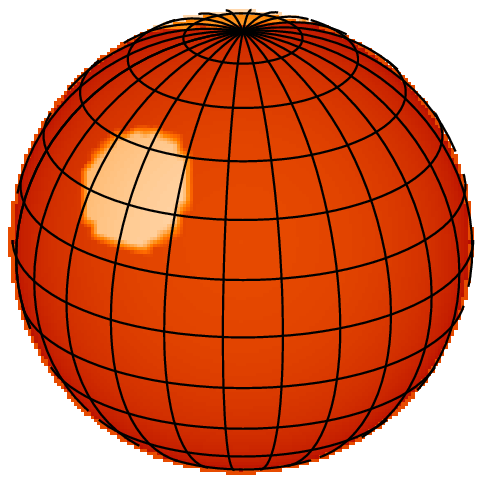}	
\hspace{0.4cm}
\includegraphics[width=2.5cm]{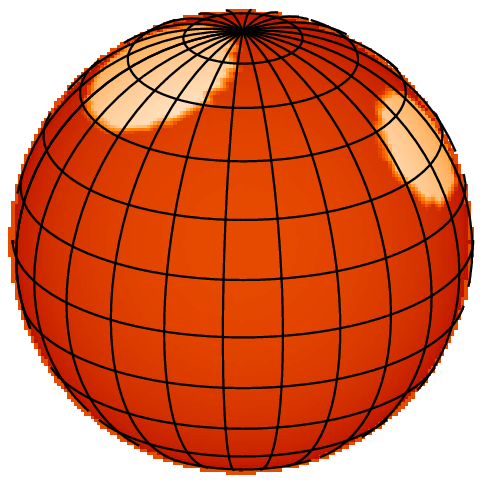}	
\includegraphics[width=2.5cm]{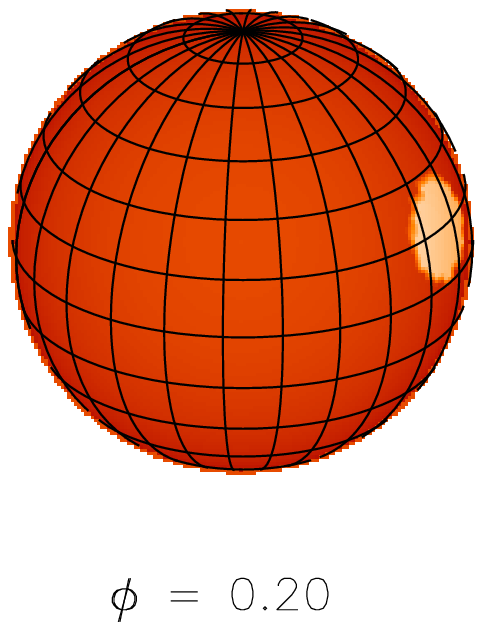}	
\hspace{0.4cm}
\includegraphics[width=2.5cm]{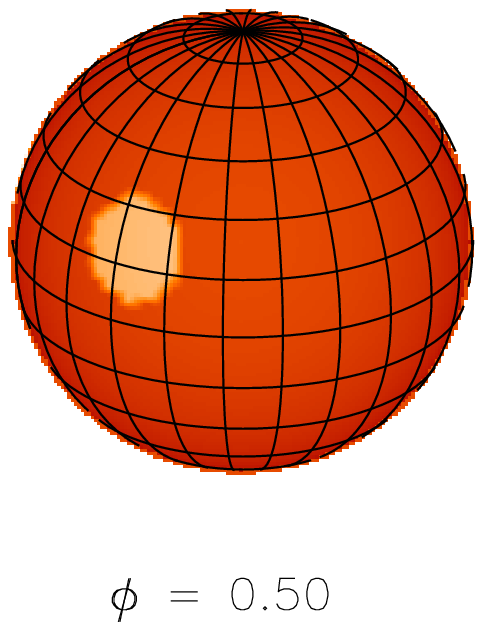}	
\hspace{0.4cm}
\includegraphics[width=2.5cm]{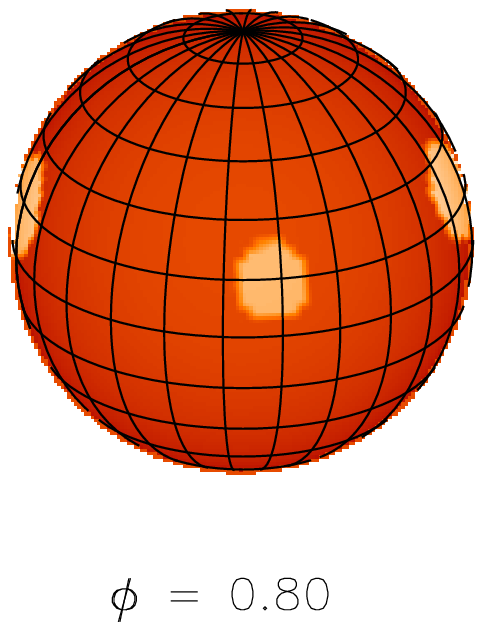}	
\includegraphics[width=2.5cm]{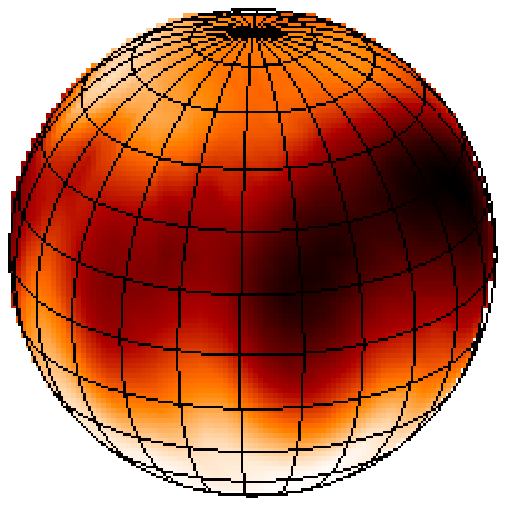}	
\hspace{0.4cm}
\includegraphics[width=2.5cm]{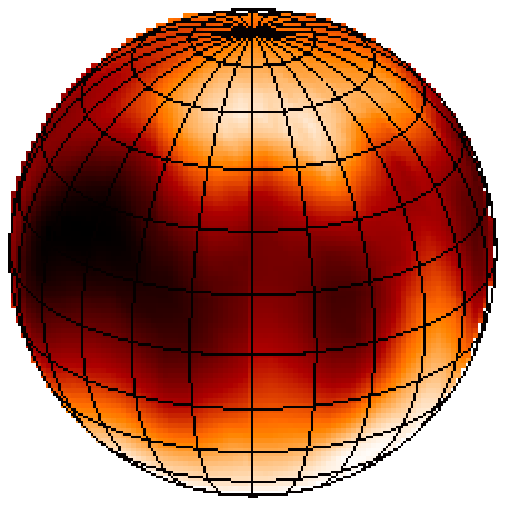}	
\hspace{0.4cm}
\includegraphics[width=2.5cm]{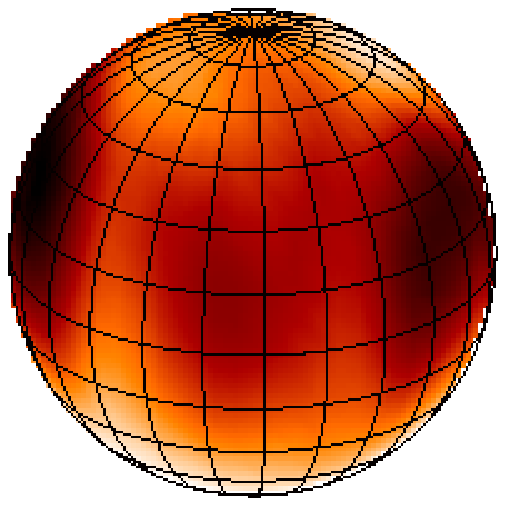}	
\end{center}
\caption{Schematic representation of the surface features of LQ~Hya in the chromosphere as reconstructed by the 
2-plage {\it (top)} and by the 3-plage models {\it (middle)} and in the photosphere {\it (bottom)}. In these maps,
the star is rotating counterclockwise.
 Dominant features are found at similar longitudes at both chromospheric and photospheric levels. }
\label{fig:palle}
\end{figure}

 The resulting plage longitudes are in good agreement with the Doppler
results in Paper\,I; i.e., photospheric minima at 0.6 and 0.9 with
the largest/coolest photospheric spots correspond to the most
luminous chromospheric phases, and vice versa, the brightest
photometric phase at 0.3 overlap with the lowest chromospheric \ha
emission, again supporting the paradigm that photospheric spots 
are physically connected with chromospheric plages
(Fig.~\ref{fig:palle}).

The synthetic \ha EW curve for the model with three plages is
plotted in the middle panel of Fig.~\ref{fig:rotmod} as a continuous
line superimposed on the data (dots), while the 2-plage
solution is displayed by a dotted line. Both of them reproduce the observed 
rotational modulation quite well. With the
same models we were able to calculate the radial velocity
shift between the \ha synthetic emission profile, resulting from the
quiet chromosphere plus the visible plages, and the photosphere. It
is clear in the bottom panel of Fig.~\ref{fig:rotmod} that the
amplitude of the theoretical $\Delta V_{\rm em}$ curves, both for a
2-plage and a 3-plage model, is very low, consistent with the values
derived from the residual profiles ($\Delta V_{\rm em}^{\rm res}$)
 and in total disagreement with the velocity shifts derived from
the raw spectra. This strongly supports the spectral synthesis 
method we used to evaluate the chromospheric emission in the \ha line.

We would like to outline the rotational modulation of other features observed
in the \ha profile, such as the peak separation, which is related to the chromospheric 
electron density (Sect.~\ref{results}) and the asymmetry of blue/red peak intensity 
(Fig.\,\ref{fig:Ne}).
The chromosphere of LQ~Hya displays a higher electron density above active regions.
The blue emission peak is stronger than the red one ($\frac{F_{\rm R}}{F_{\rm V}}< 1$)
in the ``quiet" chromosphere of LQ~Hya, while nearly equal peaks ($\frac{F_{\rm R}}{F_{\rm V}}\simeq 1$) 
tend to be observed when the most active regions are in the visible hemisphere.

Asymmetry in the peaks of an emission line with a central reversal has been frequently
observed for \ion{Ca}{ii} K line both in the Sun \citep[e.g.,][and reference therein]{Ziri88,Ding98}
and in cool stars \citep[e.g.,][and reference therein]{Montes00}.	
The blue asymmetry is more frequently observed in the solar chromosphere and is commonly 
attributed to the propagation of acoustic waves \citep[e.g.,][]{Cram76} with an upward velocity
on the order of 10 \kms\ in the layer in which the K$_2$ emission peaks are formed or a downward motion 
in the layer producing the central reversal K$_3$ \citep[e.g.,][]{Durrant76}.

\citet{Oranje83} has shown that the average \ion{Ca}{ii}~K emission profile in solar plages has a
completely different shape from the surrounding chromosphere, with the red peak slightly stronger
than the blue one. This could reflect the different physical conditions and velocity fields in active
regions compared to the quiet chromosphere.  A similar analysis cannot be made for the Sun
as a star in H$\alpha$, because this line is an absorption feature in the quiet chromosphere and
only a filling in is observed in plages.  
However, similar changes of the \ha line profile can be expected in the plages of any stars that are 
more active than the Sun. Thus, the rotational modulation of $\frac{F_{\rm R}}{F_{\rm V}}$ is consistent 
with the presence of plages in the chromosphere of LQ~Hya. 

These results suggest scaled-up versions (regarding size and brightness) 
of solar-type plages for the \ha features observed in LQ~Hya in 2000, without any 
clear evidence of strong mass motions in its chromosphere.

\begin{table}
\caption{Plage parameters for LQ~Hya in April-May 2000.}	
\label{tab:param_plages}
\begin{center}
\begin{tabular}{crcc}
\hline
\hline
\noalign{\smallskip}
   Radius  &  Lon.     &   Lat.    &  $\frac{F_{\rm plage}}{F_{\rm chrom}}$ \\
 ($\degr$) & ($\degr$) & ($\degr$) &  \\
\noalign{\smallskip}
\hline
\noalign{\smallskip}
\multicolumn{4}{c}{2 Plages Solution}\\
\hline
\noalign{\smallskip}
15.5   &  215  &  35  &  5  \\
20.0   &  353  &  66  &  5  \\
\hline
\noalign{\smallskip}
\multicolumn{4}{c}{3 Plages Solution}\\
\hline
\noalign{\smallskip}
13.0   &   210  &  20  & 5  \\
13.0   &     8  &  15  & 5  \\
9.5    &   280  &  15  & 5  \\
\hline
\hline
\end{tabular}
\end{center}
\end{table}

\begin{acknowledgements}
 We are grateful to an anonymous referee for helpful comments and suggestions.
This work has been partially supported by the Italian {\em Ministero
dell'Universit\`a e  Ricerca} (MUR), which is gratefully
acknowledged. Zs. K. is a grantee of the Bolyai J\'anos Scholarship
of the Hungarian Academy of Sciences and is also grateful to the
Hungarian Science Research Program (OTKA) for support under grants T-048961 
and K 68626. K.G.S. thanks the U.S. Kitt Peak National
Observatory for the possibility to record long time series of
stellar data with the late Coud\'e-feed telescope, now retired.
\end{acknowledgements}


\end{document}